\documentclass[useAMS,usenatbib]{mn2e}
\usepackage{epsfig}

\begin{document}

\title{Analytical solution for the structure of ADAFs}

\author[M.Shadmehri ]{Mohsen Shadmehri \thanks{E-mail: m.shadmehri@gu.ac.ir} \\
School of Physics, Faculty of Sciences, Golestan University, Gorgan 49138-15739, Iran}

\maketitle

\date{Received ______________ / Accepted _________________ }

\begin{abstract}
The standard Advection-Dominated Accretion Flow (ADAF) is studied using a set of self-similar analytical solutions in the spherical coordinates.  Our new solutions  are useful for studying ADAFs without dealing with the usual mathematical complexity. We assume the $r\varphi$ component of the stress tensor dominates and the latitudinal component of the velocity is negligible. Moreover, the fluid is incompressible and the solutions are radially self-similar. We show that our analytical solutions display most of the important properties of ADAFs which have already been obtained by  the detailed numerical solutions. According to our solutions,  the density and the pressure of the flow decreases from the equator to the polar regions and this reduction depends on the amount of the advected energy. We also show analytically that an ADAF tends to a quasi-spherical configuration  as more energy is advected with the radial flow.
\end{abstract}

\section{Introduction}
 Various  theoretical models  have been proposed to understand accreting systems over at least the last four decades \citep*[e.g.,][]{shakura73,Ichi,anderson,abra,narayan94,chen,narayan97,bland,Igu}. In these models,  mechanisms of  the energy transport in an accreting system and its radiative efficiency are among the most important physical factors.  The standard model of the accretion discs \citep*{shakura73} is successful in explaining the spectrum of some of the accreting systems such as discs around young stars \citep*[for a good review, e.g.,][]{hart}. But  radiatively inefficient accretion flows have also been proposed \citep*[e.g.,][]{Ichi,narayan94} to explain other astronomical objects like discs in active galactic nuclei (AGNs). This type of the accretion flows is generally hot, because the heat due to the turbulence is advected with the flow instead of radiating out of the system. Although the idea of such flows originally proposed by \cite{Ichi}, the first analytical description of Advection-Dominated Accretion Flows (ADAFs) has been presented by \cite{narayan94} using a set of height-integrated similarity solutions. ADAFs are generally hot and geometrically thick and their rotational velocity is sub-Keplerian due to a non-negligible effect of the gradient of pressure in the radial direction. On the other hand, when there is efficient cooling, the rotational profile  tends to the Keplerian profile which is similar to the standard disc configuration.

 Type of the accretion flow can be classified by the ratio $\dot{M}/\dot{M}_{\rm E}$, where $\dot{M}$ is the mass-accretion rate and $\dot{M}_{\rm E}$ is  the Eddington accretion rate. Structure of the disc  is described by the standard accretion model \citep*{shakura73}, if  we have $\dot{M}\leq \dot{M}_{\rm E}$. For the accretion rate much smaller than the Eddington accretion rate (i.e., $\dot{M}\ll \dot{M}_{\rm E}$), on the other hand, the optically-thin ADAF solutions are appropriate for describing the flow \citep*[e.g.,][]{Ichi,narayan94}. The disc would be optically-thick  or slim disc \citep[e.g.,][]{abra} if the accretion rate becomes larger than the Eddington rate ($\dot{M}\gg \dot{M}_{\rm E}$). Hot accretion seems to be applicable in describing properties of accretion flows around black holes in X-ray binaries and AGNs \citep*[e.g.,][]{greene,kording,ludwig}. Also, one of the best astronomical objects for the study of hot  accretion flows is Sgr ${\rm A}^{*}$ \citep[e.g.,][]{falcke97,falcke,bower,yusef}.

For analyzing steady-state structure of ADAFs, one can start with  the standard hydrodynamical equations in  cylindrical or spherical coordinates. Original study of \cite{narayan94} uses a height-integrated version of the basic equations in the cylindrical coordinates. Subsequent studies extensively extended these solutions by considering various physical ingredients like magnetic field, thermal conduction or even outflows \citep*[e.g.,][]{zhang,bu}. Although these solutions are fully analytical, one should note that  the three dimensional structure of ADAFs can not be described using these vertically {\it averaged} solutions properly, because  height-integration is a poor approximation when the flow is geometrically thick. These points motivated \cite{narayan95} (hereafter NY95) to re-analyze steady-sate structure of ADAFs in  spherical polar coordinates where the  central mass  is located at the center of the system. These solutions have also been extended by many authors over the years \citep*[e.g.,][]{tanaka,Xue,bu,jiao}.

But {\it none} of the previous works have reported a fully analytical solutions for the structure of ADAFs in  spherical polar coordinates to the best of our knowledge. In this paper, we report a fully analytical solution in the spherical coordinates. Our analytical solutions, despite their simplicity, display most of properties of the previous numerical or semi-analytical steady solutions of ADAFs in the spherical system. In the next section, basic equations of a standard ADAF model in the spherical coordinates are presented. We obtain similarity solutions in section 3 and their properties are explored. We conclude with a summary of our results in the final section.

\section{General formulation}
Our basic equations are the standard hydrodynamic   equations in the spherical coordinates $(r,\theta,\varphi)$ where the central object with mass $M$ is at its center. Temporal variation of the physical quantities is not considered which means the flow is steady state. The flow is assumed to be axisymmetric and the radial and the rotational components of the velocity are  considered.  But the latitudinal component of the velocity is assumed to be zero, i.e. $v_{\rm\theta}=0$. However, this assumption is relaxed by some authors who are interested in investigating the steady-state structure of ADAFs with outflows \cite[e.g.,][]{jiao}. We do not consider outflows, and so, the continuity equation for our incompressible flow becomes
\begin{equation}\label{eq:con}
\frac{1}{r^2}\frac{\partial}{\partial r} (r^2 \rho v_r )  =0,
\end{equation}
where $\rho(r,\theta)$ and $v_{\rm r}(r,\theta)$ are the density and the radial component of velocity.

 Now, we can write three components of the momentum equation. Assuming that the $r\phi$ component of the viscosity stress tensor to be dominant is a key assumption which greatly simplifies the equations by reducing the number of terms. Doing so, the viscous terms do not appear in the radial and the latitudinal components of the equation of motion, but viscous term has a vital role in the azimuthal component of the momentum equation \cite[also see,][]{jiao}. This simplification not only reduces the order of the differential equations, but also brings down   number of the necessary boundary conditions. Moreover, the $\alpha$ viscosity prescription is used. The shearing box magnetohydrodynamics simulations have also shown that the vertically averaged stress is proportional to the vertically averaged total thermal pressure \cite[e.g.,][]{hiro}. Thus, the components of equation of motion become
\begin{equation}
v_r \frac{\partial v_r}{\partial r}  - \frac{v_{\phi}^2}{r} = - \frac{GM}{r^2} - \frac{1}{\rho} \frac{\partial p}{\partial r},
\end{equation}
\begin{equation}
 \frac{1}{\rho r} \frac{\partial p}{\partial\theta}- \frac{v_{\phi}^2}{r} \cot\theta =0,
\end{equation}
\begin{equation}
v_r \frac{\partial v_{\phi}}{\partial r} +  \frac{v_{\phi} v_{r}}{r}  = \frac{1}{\rho r^3} \frac{\partial}{\partial r} (r^3 t_{\rm r\phi}),
\end{equation}
where  $p(r,\theta)$ and $v_{\rm\varphi}(r,\theta )$ are  pressure and the rotational velocity of the flow, respectively. Here, $t_{r\phi}$ is the $r\phi$ component of the viscosity tensor and we prescribe it based on the $\alpha$ prescription, i.e. $t_{\rm r\phi}=-\alpha p$.

 The energy equation reduces to a simple form, if we assume the $r\varphi$ component of the stress tensor is dominant. The energy equation is written as  \cite[also see,][]{jiao}
\begin{equation}
\rho v_r \frac{\partial e}{\partial r}   - \frac{p}{\rho }   v_r \frac{\partial\rho}{\partial r}  = f t_{r\phi} r \frac{\partial}{\partial r} (\frac{v_\phi }{r}),
\end{equation}
where $f$ is the advective factor and $e$ is the internal energy of the gas,
\begin{equation}\label{eq:e}
\rho e = \frac{p}{\gamma -1},
\end{equation}
where $\gamma$ is the heat capacity ratio. We also assume that both the input parameters $f$ and $\gamma$ are constant.

 Thus, equations (\ref{eq:con})-(\ref{eq:e}) constitute our basic equations of the model to be solved subject to the appropriate boundary conditions. Most of the previous semi-analytical studies generally construct radially self-similar solutions where the structure equations are solved numerically by integrating over the polar angle. We also assume radial self-similarity, but the polar angle parts are obtained analytically. Moreover, the fluid is incompressible, the $r\varphi$ component of the stress tensor dominates and $v_{\rm\theta}=0$. In the next section, we obtain our analytical solutions based on these basic assumptions.

\section{analysis}
%
%
%\subsection{self-similar solutions}
%
We introduce the following self-similar solutions:
\begin{equation}
\rho (r, \theta ) = \rho(\theta ) r^{-3/2},
\end{equation}
\begin{equation}
v_{r} (r,\theta ) = v_{r}(\theta ) \sqrt{GM/r},
\end{equation}
\begin{equation}
v_{\phi } (r,\theta ) = v_{\phi }(\theta ) \sqrt{GM/r},
\end{equation}
\begin{equation}
p(r,\theta ) = p(\theta ) GM r^{-5/2}.
\end{equation}

Now, we can substitute self-similar solutions into the above basic equations. Thus,
\begin{equation}\label{eq:r}
5p(\theta ) + \rho(\theta ) [v_{r}(\theta )^{2}-2+2 v_{\phi}(\theta )^2 ]  =0,
\end{equation}
\begin{equation}{\label{eq:theta}}
\frac{dp(\theta )}{d\theta} -   \rho (\theta ) v_{\phi}(\theta )^2  \cot\theta =0,
\end{equation}
\begin{equation}\label{eq:phi}
\alpha p(\theta ) + \rho (\theta ) v_{r}(\theta ) v_{\phi }(\theta ) =0,
\end{equation}
\begin{equation}\label{eq:energy}
(3\gamma - 5) v_{r} (\theta ) - 3\alpha f (\gamma -1) v_{\phi}(\theta ) = 0.
\end{equation}

As we show below, these differential and algebraic equations are integrable. As  boundary conditions, we assume the flow is symmetric with respect to the equatorial plane $\theta =\pi/2$ and it is sufficient to have  one of the physical quantities, say density. The rest of the quantities at the equatorial plane are obtained from the equations. We obtain density at the equatorial plane from the accretion rate (see below).

%\subsection{Nonmagnetic solutions}
%
We can now obtain the angular part of  solutions analytically. To our knowledge, all previous self-similar solutions for ADAFs in spherical coordinates  are not fully analytical. But our simple analytical solutions represent some of the basic features of ADAFs clearly.  From energy equation (\ref{eq:energy}),  we have
\begin{equation}\label{eq:non-vr}
v_{r}(\theta ) = - \frac{\alpha }{\epsilon ' } v_{\phi }(\theta ),
\end{equation}
where $\epsilon ' = \epsilon / f$ and $\epsilon = (5/3 - \gamma)/(\gamma -1)$.
Substituting the above equation for $v_{r}(\theta )$ into the equation (\ref{eq:phi}), we obtain
\begin{equation}\label{eq:non-p}
p(\theta ) = \frac{1}{\epsilon '} \rho (\theta ) v_{\phi }(\theta )^2.
\end{equation}
Using the above equation and equation (\ref{eq:non-vr}), the rotational velocity is obtained from (\ref{eq:r}), i.e.
\begin{equation}\label{eq:vphi}
v_{\phi } (\theta ) =  \frac{\sqrt{2}\epsilon ' }{g(\alpha , \epsilon ')},
\end{equation}
and equation (\ref{eq:non-vr}) gives
\begin{equation}\label{eq:vrrr}
v_{\rm r}(\theta ) =  - \frac{\sqrt{2}\alpha }{g(\alpha , \epsilon ')}.
\end{equation}
where $g(\alpha , \epsilon ') = \sqrt{\alpha^2 + 5 \epsilon ' + 2 \epsilon '^2}$.
Thus, both the radial and the rotational velocities have no dependence on the polar angle $\theta$.

Having equation (\ref{eq:non-p}),  we can integrate equation (\ref{eq:theta}) analytically, i.e.
\begin{equation}
p(\theta ) = p(\frac{\pi}{2}) (\sin\theta)^{\epsilon '},
\end{equation}
and then
\begin{equation}
\rho(\theta ) = \rho (\frac{\pi}{2})(\sin\theta )^{\epsilon '},
\end{equation}
where $p(\pi /2) = (2\epsilon ' /g)\rho(\pi /2)$.

The mass accretion rate $\dot{M}$ is written as
\begin{equation}
\dot{M}=-\int 2\pi r^2 \sin\theta \rho(r,\theta) v_{\rm r}(r,\theta) d\theta .
\end{equation}
We determine density at the equatorial plane by assuming the accretion rate is fixed. By substituting  our solutions into the above equation, we obtain
\begin{equation}
\rho (\frac{\pi}{2})= \frac{\dot{m}}{[-v_{\rm r}(\pi/2)] I(\epsilon ')},
\end{equation}
where the non-dimensional accretion rate is $\dot{m}=\dot{M}/(2\pi \sqrt{GM})$ and
\begin{equation}
I(\epsilon ')=\int_{0}^{\pi} (\sin\theta)^{1+\epsilon '} d\theta =\sqrt{\pi} \frac{ \Gamma (1+\frac{\epsilon '}{2})}{\Gamma (\frac{3}{2}+\frac{\epsilon '}{2})},
\end{equation}
where $\Gamma$ is the standard Gamma function. Therefore,
\begin{equation}
\rho (\frac{\pi}{2})= \frac{g(\alpha , \epsilon ')}{\sqrt{2\pi} \alpha } \frac{\Gamma (\frac{3}{2}+\frac{\epsilon '}{2})}{\Gamma (1+\frac{\epsilon '}{2})} \dot{m},
\end{equation}
and
\begin{equation}
p(\frac{\pi}{2}) = \sqrt{\frac{2}{\pi}} \frac{\epsilon '}{\alpha } \frac{\Gamma (\frac{3}{2}+\frac{\epsilon '}{2})}{\Gamma (1+\frac{\epsilon '}{2})} \dot{m}.
\end{equation}
\begin{figure}
%\vspace{-65pt}
\epsfig{figure=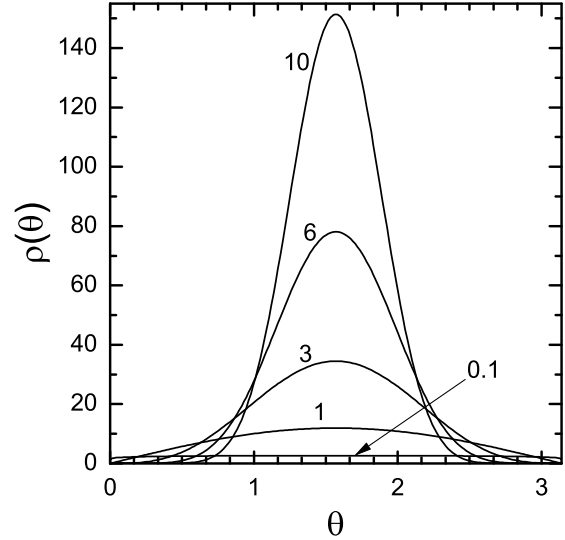,angle=0,scale=0.55}
\caption{Profile of the density versus polar angle $\theta$ for $\dot{m}=1$ and different values of $\epsilon '$. Numbers are the corresponding values of $\epsilon '$.}
\label{fig:f1}
\end{figure}
\begin{figure}
\vspace{-70pt}
\epsfig{figure=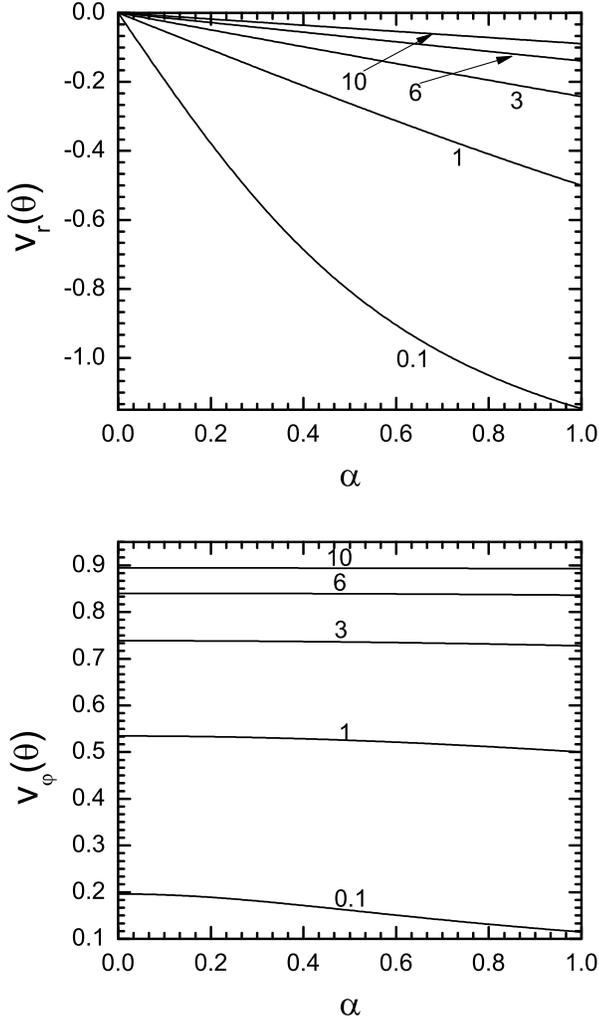,angle=0,scale=0.55}
\caption{Variation of the radial velocity ({\it top}) and the rotational velocity ({\it bottom}) versus the coefficient of viscosity $\alpha$ for different values of $\epsilon '$. Each curve is labeled by the corresponding value of $\epsilon '$.}
\label{fig:f2}
\end{figure}

Figure \ref{fig:f1} shows profile of the density versus the polar angle for $\dot{m}=1$ and different values of $\epsilon '$. Each curve is labeled by the corresponding value of $\epsilon '$. For a fixed $\gamma$, a larger $\epsilon '$ implies less advected energy with the flow. For small values of $\epsilon '$ which means the flow is fully advecteive, we see that the density has a little variation from the pole to the equator. Thus, the solution corresponds to a nearly spherical accretion flow. But as the value of $\epsilon '$ increases, the mass distribution concentrates more around the equatorial regions so that the density contrast between $\rho (0)$ and $\rho (\pi /2)$  is enhanced significantly. A case with a large value of $\epsilon '$ resembles to a standard thin disc configuration where there is significant cooling. Behavior of the density distribution based on our analytical solution is very similar to what has been shown by NY95 in their Figure 1.

 We also found that both $v_{\rm r}(\theta)$ and $v_{\rm\varphi}(\theta )$ are independent of the polar angle. The $\theta -$independence of the components of the velocity arise almost by construction, i.e. from equations (\ref{eq:vphi}) and (\ref{eq:vrrr}). This is a major qualitative difference in comparison to the NY95 solutions where the radial velocity is zero at the poles and increases monotonically toward the equator. NY95 showed that  variation of $v_{\rm r}(\theta)$ with the polar angle becomes less significant as the value of $\epsilon '$ increases. Some authors also found outflows at the poles \cite[see e.g.,][]{tanaka,jiao}. As for the rotational velocity and the temperature there is no polar dependence according to our new solutions. In the NY95 solutions, the square of the sound speed can vary by almost an order of magnitude between the pole and the equator. In contrary to these qualitative differences, we think, our $\theta$-independent values for the radial and the rotational velocities and the sound speed are consistent with the polar-angle average of the NY95 solutions. Although our new solutions do not contain polar angle dependence, these solutions recover these physical quantities in an angle-averaged sense.

Now, we explore variations of the radial and the rotational velocities with the input parameters.  In Fig. \ref{fig:f2}, variations of the radial velocity (top) and the rotational velocity (bottom) versus $\alpha$ for different values of $\epsilon '$ are displayed. For a fixed value of $\alpha$, as more dissipated turbulent energy is advected with the flow, deviation of the rotational velocity from the Keplerian profile becomes more significant. We also showed that the radial velocity decreases with increasing $\epsilon '$. Moreover, rotational velocity has   little dependence on the value of $\alpha$ unless the flow becomes fully advective. But the radial velocity significantly increases with $\alpha$, though its dependence on the value of viscosity coefficient becomes less significant as more energy radiates out of the system. Therefore, our solutions display most of the properties of the components of the velocity in an angle-average sense which have already been obtained by NY95.

 We can also calculate the Bernoulli parameter $Be$ corresponding to our solutions. If this parameter is positive, then it means we may have outflows.  Based on this hypothesis, NY95 calculated the Bernoulli parameter for their solutions and found ranges of the input parameters within which  parameter $Be$ is positive. Subsequent studies extensively studied  outflows from ADAFs using similarity methods \citep*[e.g.,][]{Xue,tanaka,jiao}. However, \cite{tanaka} found in their solutions no relation between outflows and the Bernoulli parameter. We note that the outflow found in \cite{tanaka} is driven by thermal conduction (as opposed to the way the viscous flow equations are solved). Here, we calculate $Be$ of our solutions to see if the profile of this parameter is similar to what has been obtained by NY95. The non-dimensional Bernoulli parameter $b$ is written as
\begin{equation}
b=\frac{Be}{\Omega_{\rm K}r^2}=\frac{1}{2}v_{\rm r}^2 + \frac{1}{2} (v_{\rm\varphi} \sin\theta)^2 - 1 + \frac{\gamma}{\gamma -1} c_{\rm s}^2 ,
\end{equation}
where $c_{\rm s}$ is the sound speed and $\Omega_{\rm K}$ is the Keplerian angular velocity. Upon substituting our solutions into the above equation, we obtain
\begin{equation}
b(\theta ) = \left (\frac{\epsilon '}{g} \right )^{2}\left ( \sin^2\theta + 3f -2 \right ).
\end{equation}

Clearly, the parameter $b(\theta)$ is positive to all $\theta$ for $f>2/3 \approx 0.66$. But this critical value of the advection parameter has been found by NY95 as $f\approx 0.446$.  But our Bernoulli parameter is qualitatively different from the behavior of the NY95 solutions. More specifically, the new solutions have $b(\theta )$ increasing monotonically toward the equator which is exactly opposite of the behavior of the NY95 solutions. In the new solutions, radial and the rotational velocity and the sound speed are all constant with respect to the polar angle because of our simplifying assumptions. In the previous solutions, however, all of these depended on the polar angle. Since the new solutions give values for $v_{\rm r}$, $v_{\rm\varphi}$ and $c_{\rm s}$ that are consistent with the polar angle-averaged values of the previous solutions, the computation of the Bernoulli parameter seems to be only valid in angle-averaged magnitude.

\section{Conclusions}

We reported a set of self-similar {\it analytical} solutions for the structure of ADAFs in  spherical polar coordinates. Although our solutions  are obtained  without solving partial differential equations numerically,  their physical properties are  similar to the previous known ADAF solutions.  Most of the previous similarity solutions for the steady state structure of ADAFs in the spherical coordinates are obtained from solving a set of differential equations subject to the suitable boundary conditions at the pole and the equator which seem to be a challenging problem. But our analytical solutions could be useful for the authors wishing to study properties of ADAFs without struggling with those numerical difficulties.

 Because of our simplifying assumptions, however, physical quantities such as components of the velocity and the sound speed are independent of the polar angle. But in an angle-average sense these solutions are consistent with previous solutions despite of missing their polar angle dependence entirely. Our similarity solutions are based on a few  assumptions such as neglecting  all components of the stress tensor except the $r\varphi$ component. This key hypothesis greatly reduces the number of terms in the momentum and the energy equations. The solutions are also parameterized by a fixed accretion rate. This enabled us to calculate the density and the pressure at the equator analytically in terms of non-dimensional accretion rate, coefficient of viscosity and the amount of the advected energy.

We also showed analytically that the geometrical thickness of the flow sensitively depends on the amount of the advected energy. Note that density and the pressure distributions are obtained without height-integration procedure.  As more energy is advected with the radial flow, not only the temperature increases, but also the density distribution tends to a quasi-spherical configuration. Although this typical behaviour of ADAFs has already been discussed through detailed numerical approaches, our model confirms this behavior by an analytical solution. Moreover,  the advantage of having analytic solutions for the structure of ADAFs allows one to use these solutions for the modeling astrophysical systems where properties of the ADAFs are among the input parameters.

\section*{Acknowledgments}
I am grateful to the anonymous referee whose detailed and careful comments helped to improve the quality of this paper.

\bibliographystyle{mn2e}
\bibliography{referenceKH}

%\appendix

%
%
%
\end{document}